\documentclass[final,twocolumn,5p,authoryear]{elsarticle}

\usepackage{amssymb}

\usepackage{epsfig}         
\usepackage{color}          
\usepackage{times}          
\usepackage{amsmath}        
\usepackage{graphicx}

\journal{Journal of High Energy Astrophysics}

\begin{document}

\begin{frontmatter}



\title{Theoretical model of hydrodynamic jet formation from
accretion disks with turbulent viscosity}


\author[label1,label2]{E. Arshilava}
\address[label1]{Department of Physics, Faculty of Exact and Natural Sciences,
     Javakhishvili Tbilisi State University (TSU), Tbilisi 0179, Georgia}
\address[label2]{Department of Physics and Astronomy, Heidelberg University, 69120, Heidelberg, Germany}
\author[label1,label3]{M. Gogilashvili}
\address[label3]{Department of Physics, Florida State University, Tallahassee, FL 32306, USA}
\author[label1,label3]{V. Loladze}
\author[label1]{I. Jokhadze}
\author[label1,label5]{B. Modrekiladze}
\address[label5]{Department of Physics, Carnegie Mellon University, 5000 Forbes Ave., Pittsburgh, PA 15213, USA}
\author[label1,label4]{N. L. Shatashvili}
\address[label4]{TSU Andronikashvili Institute of Physics, TSU, Tbilisi 0177, Georgia}
\author[label1,label5,label6]{A. G. Tevzadze}
\address[label6]{Abastumani Astrophysical Observatory, Ilia State University, 3/5 Cholokashvili Ave., Tbilisi 0162, Georgia}

\begin{abstract}
We develop the theoretical model for the analytic description
of hydrodynamic jets from protostellar disks employing the
Beltrami-Bernoulli flow configuration of disk-jet structure.
For this purpose we extend the standard turbulent viscosity
prescription and derive several classes of analytic
solutions using the flow parametrization in self-similar variables. Derived
solutions describe the disk-jet structure, where for the
first time jet properties are analytically linked with the properties of
the accretion disk flow. The ratio of the jet ejection and disk accretion
velocities is controlled by the turbulence parameter, while the ejection
velocity increases with the decrease of local sound velocity and the jet
launching radius. Derived solutions can be used to analyze the
astrophysical jets from protostellar accretion disks and
link the properties of outflows with the local observational
properties of accretion disk flows.
\end{abstract}


\end{frontmatter}




\section{Introduction}
\label{Intro}

Jets streaming from young stellar objects (YSOs) are spectacular
manifestations of the star formation process. These outflows, believed to be
powered by protostellar accretion disk, carry away the matter, energy and
angular momentum of the accreting matter, thus promoting the development of a
protostar. Properties of these disk-jet systems are inferred from the
observations of Herbig-Haro (HH) objects, T-Tauri and Herbig Ae/Be stars,
protostellar and protoplanetary disks.

Collimated bipolar outflows from YSO are known to be parsec-scale,
non-relativistic and supersonic by nature. The radial velocity of the jet-flow
varies from 20 km/s up to 450 km/s for HH jets (see
\citealt{Hartigan2011,Plunkett2015,Podio2016,Jhan2016,Bjerkeli2016,Reitar2017}).
These jets are launched in the vicinity of a protostar, as close as 0.03au
(see \citealt{Lee2017}). A vast number of collimated jets are detected in
molecular clouds pointing to the embedded protostars. Surveys of these jets
(see, e.g., \citealt{Ioannidis2012,Smith2014,Zhang2014}) provide a large
unbiased observational data that can be used to test the theoretical models of
the disk-jet structures. Today we know that morphology, sizes and velocities
of the jet-outflows can be used to estimate the mass,
luminosity and/or age of the YSOs (see \citealt{Bally2016} and references
therein).

Another class of wide-angle hydrodynamic outflows from HH
objects decelerate either with increasing the angle away from the central axis
of the flow, or increasing distance along axis (see \citealt{Lizano1988}). It
is known that these hydrodynamic outflows of neutral atoms are
intrinsically linked with the process of star formation (see, e.g.,
\citealt{Ruden1990}, \citealt{Shu1991}).

The main mass/energy source of the ejecta is the disk-flow material/energy
released through the accretion process. Hence, jet velocities are
intrinsically correlated with the accretion rates calling for the unified
treatment of the disk-jet system dynamics. The central object dynamics may
play the additional role in the formation of relativistic outflows/jets
[\citealt{Livio}-- powerful jets are produced by systems in which on top of an
accretion disk threaded by a vertical field, there exists an additional source
of energy/wind, possibly associated with the central object (for example,
stellar wind from porotostar may accelerate YSO jets, as estimated by
\citealt{Fereira1997,Fereira}]. In (\citealt{SY2011}) it was shown, that there
exists a general principle that dictates a marked similarity in macroscopic
disk-jet geometry despite the huge variety of the scaling parameters such as
Lorentz factor, Reynolds number, Lundquist number, ionization fractions, etc.,
characterizing different systems.

A class of the hydrodynamic jet solution from viscous accretion flows has been
studied by \citealt{Scott1987}. Self-similar solutions of the model give
velocity field being inverse proportional to the radial
distance from the center. Matching the profiles of circular jets
(\citealt{Squire1951}) this solution deviates from the Keplerian profile and
can be effectively used outside the accretion disk area. Another
solution to the formation of hydrodynamic jets has been studied by
\citealt{Hernandez2014} showing that under certain conditions
hydrodynamic accretion disk can develop instabilities. Using
perturbation analysis authors have shown that developed
instability may lead to the formation of pair of bipolar
jet-outflows. Instability mechanism is unlikely to support a
steady disk-jet structure, and can possibly be used in the description of
short time outbursts from the hydrodynamic accretion flows in
the vicinity of central object. Recently \citealt{Clarke2016} have used
a self-similar approach for the description of the axisymmetric
hydrodynamic outflows from hydrodynamic accretion disks. Authors have shown
isothermal disk-wind structure that can be developed for the power low density
profile that well matches the results of numerical simulations in the wind
area.

Since the viscous accretion disk model was proposed (\citealt{Shakura1973}),
with an observed high ejection efficiency, it became natural to assume that
jets are driven magnetically from an accretion disk --- when magnetic field is
advected inwards by accreting material or/and generated locally by some
mechanism, the centrifugal force due to rotation may boost the jet along the
magnetic field lines up to a super-Alfv\'enic speed,
(\citealt{begelman3,begelman4,begelman5,bland4,shibata,shibata2,anderson2}).
Blandford \& Payne (1982) studied the magneto-centrifugal
acceleration along the magnetic field lines and demonstrated that the magnetic
field results into instability of particles at the Kepler orbit leading to the
Jet formation in the disk center (with the opening angle of the jet $\leq
30^{o}$). They were the first to show the braking of matter in the azimuthal
direction inside the disk and the outflow acceleration above the disk surface
guided by the poloidal magnetic field components. Toroidal components of the
magnetic field then collimate the flow. In the case of the fully turbulent
Keplerian disk the poloidal magnetic field tends to drift outward
(\citealt{lovelace1,lubow,lovelace2}) so that its value cannot significantly
exceed the strength of the large-scale seed magnetic field. Then, according to the
hydro-magnetic models, the magnetic fields provide a natural mechanical link
between disks and jets and can account for the launching, confinement and
collimation of jets (see e.g.
(\citealt{bland,bland1,bland1-2,bland2,bland3,zanni}))
--- the angular momentum, energy and mass can be removed from the
accreting flow. The collimation is provided by the stratified thermal pressure
from an external medium while the acceleration efficiency then depends on the
pressure gradient of the medium.

In the present paper we present the results of theoretical study of the
disk-jet structure formation for YSOs based on the Beltrami Flow model
(\citealt{SY2011}) using the Turbulent Viscosity approach
(\citealt{Shakura1973}) as the main reason of accretion; disk was
assumed un-magnetized, hence, there is no pre-existed global magnetic field.
We have found the analytical conditions for disk-jet structure formation and
parameter ranges for Jet-launching and collimation for YSO Jets.

The Physical model of the problem is described in Sec 2 where we derive the
extended turbulent viscosity model and use it in the model equations for the
disk-jet flows. Using the similarity variables and variable splitting ansatz
we derive several classes of the solutions, which includes the flow
configuration of the accretion disk--ejection jet structure. We use specific
examples to illustrate the disk-jet structures and explore their properties.
The paper is summarized in Sec. 3.

\section{Physical Model}
\label{Model}

YSO disk-jet structure, according to observations, is quite a long-lived
object. Hence, the steady state solutions could well describe its behavior.
Equations governing the dynamics of the stationary viscous compressible fluid
rotating around a central gravitating object can be written as follows:
\begin{equation}
({\bf V} \cdot \nabla) {\bf V} = - \nabla H - \nabla \Phi +
{1 \over \rho} \nabla \cdot {\rm T} ~, \label{PDE1}
\end{equation}
\begin{equation}
\nabla \cdot (\rho {\bf V} ) = 0 ~, \label{PDE2}
\end{equation}
where ${\bf V}$, $\rho$ and $H$ are the velocity, density and enthalpy,
respectively; {$\Phi$ is the} gravitational potential of the central object;
it is assumed, that the gravity of disk can be ignored. Dissipative effects
are described by the viscous stress tensor \ { $T_{ik}$ and the corresponding
term in Eq.(\ref{PDE1}) is formally written as}:
\begin{equation}
\nabla \cdot {\rm T} \equiv \nabla_k \, T_{ik} = {\partial \over \partial x_k} T_{ik} ~.
\end{equation}
The barotropic equation of state (for our problem of study) was used to
calculate the enthalpy of the fluid:
\begin{equation}
\nabla H = {1 \over \rho} \nabla {\cal{P}} ~, \label{PDE3}
\end{equation}
where \ ${\cal{P}}$ \ is the thermodynamic pressure. To seek for the steady state
solutions of the disc-jet structures persisting around a central accreting
object we introduce a so-called ``ideal'' and ``reduced'' factors of the
``local'' density { following \cite{SY2011,Yoshida2012}}:
\begin{equation}
\rho = \rho_{I} \rho_{R} ~.
\end{equation}
The purpose of this separation is to separate ideal fluid and dissipative
effects { and to track the accretion effects in jet formation process}. {
Thus, in conventional ideal fluid mechanics} with zero dissipation $\rho_{R} =
1$ and $\rho_{I} = \rho$ . In such formalism we can introduce {``ideal''} and
``reduced'' momenta as:
\begin{eqnarray}
{\bf P}_{I} &=& \rho_{I} {\bf V} ~, \\
{\bf P}_{R} &=& \rho_{R} {\bf V} ~.
\end{eqnarray}
Obviously, the reduced momentum matches flow velocity in fluids with zero
dissipation.  Reformulating Eqs. (\ref{PDE1},\ref{PDE2}) in terms of the new
momenta we can derive a ``Generalized Pressure Balance equation'' in our
definitions as follows:
\begin{equation}
{\bf P}_{R} \times (\nabla \times {\bf P}_{R}) = \label{Bernoulli1}
\hskip 4 cm
\end{equation}
\begin{equation*}
= {1 \over 2} \nabla P_{R}^2 + \rho_{R}^2 \nabla \left(H + \Phi \right)
+ \frac{\rho_{R}}{\rho_{I}} \left[ {\bf P}_{R}
(\nabla \cdot {\bf P}_{I}) + \nabla \cdot {\rm T} \right] .
\end{equation*}
We will later use this equation to define the topology of the fluid and
reduced components of the disk-jet system. Firstly, the geometry of the
problem allows the assumption for the reduced momentum to obey the Beltrami
Condition implying that the Reduced Momentum is aligning along its
corresponding Generalized Vorticity (see e.g. \citealt{SY2011,Yoshida2012} and
references therein):
\begin{equation}
{\bf P}_R = \lambda \ ( \nabla \times {\bf P}_R ) ~, \label{Beltrami1}
\end{equation}
thus making the left hand side term strictly zero in the Eq.
(\ref{Bernoulli1}). Here $\lambda $ stands for the Beltrami parameter related
to the so called flow reduced momentum. Secondly, we seek for the solution of
the fluid reduced momentum that will make the last term zero on the r.h.s. of
the Eq. (\ref{Bernoulli1}), thus, fully determined by the viscosity effect:
\begin{equation}
{\bf P}_R (\nabla \cdot {\bf P}_I) + \nabla \cdot {\rm T} = 0 ~.
\label{Bernoulli21}
\end{equation}
Using such assumptions pressure balance equation (\ref{Bernoulli1}) reduces to
a ``Generalized Bernoulli Condition'' written for the reduced momentum and
the reduced density of the flow:
\begin{equation}
{1 \over 2} \nabla P_R^2 + \rho_R^2 \nabla \left(H + \Phi\right) = 0
~. \label{Bernoulli22}
\end{equation}
Hence, the stationary state of the system can be fully analyzed using Eqs.
(\ref{Beltrami1},\ref{Bernoulli21},\ref{Bernoulli22}) and the explicit form of
the viscous stress tensor related to the specific YSO conditions.
Note that the Eqs. (\ref{Bernoulli21}) and (\ref{Bernoulli22})
are valid only in the case of dissipative flow (${\rm T} \not=0$), while
inviscid Beltrami flow (${\rm T}=0$) can be described by the Eq. (\ref{Bernoulli1})
with the corresponding right hand side.

\subsection{Turbulent viscosity model}
\label{Viscosity}

The radial outward momentum transfer and consequent inward accretion is due to
the dissipative processes induced by the turbulence in the disk-jet flow.
Hence, we employ the turbulent viscosity model when the small scale
turbulence creates the anomalous dissipation that can be described by using
the $\alpha$--viscosity model introduced by \cite{Shakura1973}. In contrast
to the standard $\alpha$ model, we plan to use the effective viscosity model
both in the disk and jet as well as in the disk-jet transition areas.

We employ cylindrical coordinates \ $(r, \varphi, z)$ \ to describe the
disk-jet system. Hence, the only significant components of the viscous stress
tensor $T_{ik}$, assuming the strong azimuthal rotation, will be:
\begin{equation}
T_{r \varphi} = T_{\varphi r} = \nu_{\rm t} \rho
\left[ r {\partial \over \partial r}
\left({V_\varphi \over r} \right) +
{1 \over r} {\partial V_r \over \partial \varphi} \right] ~, \label{Trp}
\end{equation}
\begin{equation}
T_{z \varphi} = T_{\varphi z} = \nu_{\rm t} \rho
\left[ {\partial V_\varphi \over \partial z} +
{1 \over r}{\partial V_z \over \partial \varphi} \right] ~, \label{Tzp}
\end{equation}
where $\nu_{\rm t}$ is the turbulent viscosity parameter. We now split the
pressure $\cal{P} $ into the background constant component ${\cal P}_0$ and $p$
being a deviation from this background value:
\begin{equation}
{\cal{P}} = {\cal P}_0 + p ~.
\end{equation}
In this limit the turbulent stress tensor can be split into the background
constant component $\bar T_{ik}$ and smaller deviation $t_{ik}$ that varies
with spatial coordinate:
\begin{equation}
T_{ik} = \bar T_{ik} + t_{ik} ~. \label{Tik}
\end{equation}
The classical $\alpha$--viscosity model links turbulent viscosity stress
tensor to the pressure using the constant parameter $\alpha_0 $:
\begin{equation}
\bar T_{r \varphi} = \alpha_0 {\cal P}_0 ~. \label{T0ik}
\end{equation}
Assuming axisymmetric flow $V_\varphi = r \Omega_{\rm K}(r,z)$ rotating
locally with Keplerian angular velocity:
\begin{equation}
\Omega_{\rm K}^2(r,z) = {G M_\star \over (r^2 + z^2)^{3/2}} ~, \label{Omega_Kep}
\end{equation}
where $M_\star$ is the mass of the central object, we can extend the standard
turbulent viscosity model to derive:
\begin{equation}
t_{ik} \equiv \left\langle {3 \over 2} \nu_{\rm t} \rho
\Omega_{\rm K}(r,z)\right\rangle {r^2 \over r^2 + z^2} - \alpha_0 {\cal P}_0 ~.
\label{tik}
\end{equation}
Hence, in the axisymmetric case, when the azimuthal gradients in the Eqs.
(\ref{Trp},\ref{Tzp}) can be neglected, the viscous stress tensor elements can
be calculated as follows:
\begin{equation}
t_{r \varphi} = {r^2 \over r^2 + z^2} \,\beta \,p ~, \label{Tr}
\end{equation}
\begin{equation}
t_{z \varphi} = {r z \over r^2 + z^2}\, \beta  \,p ~. \label{Tz}
\end{equation}
Here $p$ can be positive/negative, corresponding to the stronger/weaker
turbulence compared to the background turbulent steady state.
Assumption of the strictly Keplerian local angular velocity of
the rotation (see Eq. (\ref{Omega_Kep})) can be justified for the rotationally
supported flow, for which the radial pressure gradients can be negligible compared to
the centrifugal force. Such situation is realized in slowly accreting flows,
where background pressure is known to vary slowly, i.e., ${\cal P}_0 = const.$
).

\subsection{Model Equations for Disk-Jet Structure}
\label{Equations}

Dictated by the geometry of observed YSOs disk-jet structure
and the continuity Eq. (\ref{PDE2}), let us expand the flow velocity
using the axisymmetric stream function $\psi$ and
local Keplerian azimuthal circulation, as follows:
\begin{equation}
{\bf V} = {1 \over \rho} \left( \nabla \psi \times \nabla \varphi \right)
+ r V_\varphi \nabla \varphi ~. \label{Stream}
\end{equation}
Interestingly, the $\psi$ introduced in such a way matches the stream function
of the actual momentum ${\bf P} = \rho {\bf V}$ of the flow. Now we can
reformulate our problem in terms of the stream function $\psi$ using the
turbulent viscosity stress tensor calculated in the previous subsection.
Hence, Eq. (\ref{Bernoulli21}) takes the following form:
\begin{equation}
{V_\varphi \over r} \left( {\partial \psi \over
\partial r} {\partial \over
\partial z} \ln \rho_R - {\partial \psi \over \partial z} {\partial \over
\partial r} \ln \rho_R
 \right) = \label{Stream1}
\end{equation}
\begin{equation*}
\hskip 2cm \beta \,{r^2  \over r^2 + z^2} \left[ {\partial p \over \partial r}
+ {z \over r} {\partial p \over \partial z} + {2\beta \over r} p \right] ~,
\end{equation*}
while the generalized Bernoulli Condition (\ref{Bernoulli22}) will be reduced
to the following:
\begin{equation}
\nabla {\cal E}_m + \left(V_\varphi^2 + {(\nabla \psi)^2 \over r^2 \rho^2} \right)
\nabla \ln \rho_R = 0 ~, \label{Stream2}
\end{equation}
where we have introduced the total mechanical energy ${\cal E}_m$ as follows:
\begin{equation}
{\cal E}_m \equiv \Phi + {V_\varphi^2 \over 2} +
{(\nabla \psi)^2 \over 2 r^2 \rho^2} ~. \label{E}
\end{equation}
The system of equations describing the YSOs disk-jet structure can be closed
using the Beltrami condition (\ref{Beltrami1}) written in the stream function
representation:
\[
\nabla\times\nabla\psi\times\nabla\varphi \ + \ \nabla\times
\left[(\rho r V_{\varphi}) \nabla\varphi \right]
\ + \ \nabla\psi\times\nabla\varphi\times\nabla\ln\rho_I
\]
\begin{equation}
+ \ (\rho r V_{\varphi}) \nabla\varphi\times\nabla\ln\rho_I =
\lambda\left(\nabla\psi\times\nabla\varphi
+ (\rho r V_{\varphi}) \nabla\varphi\right) \ .
\label{beltrami2}
\end{equation}
Hence, the system is reduced to the Eqs. (\ref{Stream1},\ref{Stream2},\ref{E})
and (\ref{beltrami2}).

\subsection{Equations in the similarity variables}
\label{Similarity}

To construct the similarity solution of the system of equations
(\ref{Stream1}-\ref{beltrami2}) representing disk-jet structure we introduce
the orthogonal variables $\tau$ and $\sigma$ as follows:
\begin{eqnarray}
\sigma &=& \sqrt{r^2 + z^2} ~, \\
\tau &=& z/r ~,
\end{eqnarray}
for which $\nabla \tau \cdot \nabla \sigma = 0$. Then, the derivatives in
$r$ and $z$ variables can be expressed as:
\begin{eqnarray}
{\partial \over \partial r} &=& -{z \over r^2} {\partial \over \partial \tau} +
{r \over \sqrt{r^2 + z^2}} {\partial \over \partial \sigma} \nonumber \\
&=& - \frac{\tau\sqrt{1+\tau^2}}{\sigma}\,\frac{\partial }{\partial \tau}
+\frac{1}{\sqrt{1+\tau^2}}\,\frac{\partial }{\partial \sigma} \ ,
\end{eqnarray}
\begin{eqnarray}
{\partial \over \partial z} &=& {1 \over r} {\partial \over \partial \tau} +
{z \over \sqrt{r^2 + z^2}} {\partial \over \partial \sigma} \nonumber \\
&=& \frac{\sqrt{1+\tau^2}}{\sigma}\,\frac{\partial }{\partial \tau}
+\frac{\tau}{\sqrt{1+\tau^2}}\,\frac{\partial }{\partial \sigma} \ .
\end{eqnarray}
The first obvious result of such coordinates is that the Gravitational
Potential depends only on one variable: $\Phi=\Phi(\sigma) = -
\Omega_0^2/\sigma$. While the natural representation of
velocity given by (\ref{Stream}) allows the simplifying assumption
for the stream function to be dependent only on the $\tau$ variable:
\begin{equation}
\psi = \psi(\tau) ~,
\end{equation}
thus, making it possible to separate the variables in the solution. Then, the
Eq. (\ref{Stream1}) will take the following form in the similarity variables:
\begin{equation}
V_\varphi {1 + \tau^2 \over \sigma^2} {\partial \psi \over \partial \tau}
{\partial \over \partial \sigma} \ln \rho_R  =
\left( {\partial \over \partial \sigma} + {2 \over \sigma} \right)
{\beta\, p \over 1+\tau^2} ~,
\label{Sim1}
\end{equation}
The three components $r, \varphi, z$ of the Beltrami conditions
(\ref{beltrami2}) in the new variables yield the following three equations,
respectively:

\begin{equation}
\left(\frac{\tau\sigma^2}{(1+\tau^2)^{3/2}} {\partial \over \partial \sigma}
+ {\partial \over \partial \tau} \right)\ \ln \left(\frac{\sigma \rho_R V_{\varphi} }
{(1+\tau^2)^{1/2}} \right)
=\frac{\lambda}{\rho V_{\varphi}}{\partial \psi \over \partial \tau} ~,
\label{Simr}
\end{equation}

\begin{equation}
{\partial^2 \psi \over \partial \tau^2} + \left( {3 \tau \over 1+ \tau^2} -
{\partial \over \partial \tau} \ln \rho_I \right)
{\partial \psi \over \partial \tau}  =
- \lambda {\sigma^3 \over (1+\tau^2)^{5/2}} \rho V_\varphi ~,
\label{Sim4}
\end{equation}

\begin{equation}
\left( - \frac{\sigma^2}{\tau(1+\tau^2)^{3/2}}{\partial \over \partial \sigma}
+{\partial \over \partial \tau} \right)
\ln \left(\frac{\sigma \rho_R V_{\varphi}}{(1+\tau^2)^{1/2}}  \right)
=\frac{\lambda}{\rho V_{\varphi}} {\partial \psi \over \partial \tau} ~.
\label{Simz}
\end{equation}
Equations (\ref{Simr}) and (\ref{Simz}), after some straightforward algebra,
lead to following elegant equations:
\begin{equation}
{\partial \over \partial \sigma} \left( {\sigma \over (1+\tau^2)^{1/2}}
\rho V_\varphi \right) =  0 ~, \label{Sim2}
\end{equation}
\begin{equation}
{\partial \over \partial \tau} \ln \left( {\sigma \over (1+\tau^2)^{1/2}}
\rho_R V_\varphi \right) =
\lambda  {(1+\tau^2)^{1/2} \over \sigma \rho V_\varphi}
{\partial \psi \over \partial \tau}  ~, \label{Sim3}
\end{equation}
while $r$ and $z$ components of the Generalized Bernoulli Condition
(\ref{Bernoulli22}) will give the following two equations, respectively:
\begin{equation}
{1 \over \rho} {\partial p \over \partial \sigma} +
{\partial {\cal E}_m \over \partial \sigma} + 2\left( {\cal E}_m - \Phi \right)
{\partial \over \partial \sigma} \ln \rho_R = 0 ~, \label{Sim5}
\end{equation}
\begin{equation}
{1 \over \rho} {\partial p \over \partial \tau} +
{\partial {\cal E}_m \over \partial \tau} + 2\left( {\cal E}_m - \Phi \right)
{\partial \over \partial \tau} \ln \rho_R = 0 ~, \label{Sim6}
\end{equation}
where the total mechanical energy of the system ${\cal E}_m$
[defined by Eq. (\ref{E})] is now expressed in similarity variables ($\sigma, \tau $).
Hence, in our new framework, the physical system  of interest
can be analyzed using Eqs. (\ref{Sim1},\ref{Sim4},\ref{Sim2}-\ref{Sim6}).

\subsection{Variable splitting ansatz}
\label{Splitting}

We seek the solution of the system assuming that describing physical variables
can be factorized using the similarity variables $(\sigma, \tau $) introduced in previous
subsection:
\begin{eqnarray}
p(\sigma,\tau) &=& p_1(\sigma) p_2(\tau) ~, \nonumber \\
\rho_{\rm I}(\sigma,\tau)  &=& \rho_{I1}(\sigma) \rho_{I2}(\tau) ~, \\
\rho_{\rm R}(\sigma,\tau)  &=& \rho_{R1}(\sigma) \rho_{R2}(\tau) ~,\nonumber \\
\lambda(\sigma,\tau)  &=& \lambda_1(\sigma) \lambda_2(\tau) ~. \nonumber
\label{Ansatz}
\end{eqnarray}
The azimuthal velocity can be calculated using the Keplerian rotation velocity
$V_\varphi(\sigma,\tau) = V_{\rm Kep}$, where
\begin{equation}
V_{\rm Kep} =  {\sigma_0 \Omega_0 \over (1+\tau^2)^{1/2}}
\left( {\sigma \over {\sigma_0}}\right)^{-1/2} ~. \label{Kepler}
\end{equation}
Here $\Omega_0$ is the Keplerian angular velocity of the rotation at some
characteristic radius $\sigma_0$ in the disk. Applying this ansatz into Eqs.
(\ref{Sim1}-\ref{Sim6}), solving them in $\sigma$,  we find that the balance of
all terms give the following solutions in the $\sigma$ coordinate:
\begin{eqnarray}
p_1(\sigma) &=& \sigma^{-5/2} ~, \nonumber \\
\rho_{\rm I1}(\sigma) &=& \sigma^{-1} ~, \\
\rho_{\rm R1}(\sigma) &=& \sigma^{-1/2} ~, \nonumber \\
\lambda_1(\sigma) &=& \sigma^{-1} ~. \nonumber
\label{Ansatz-1}
\end{eqnarray}
Having derived the radial profiles of the solution we can now
reduce Eqs. (\ref{Sim1}-\ref{Sim6}) to the system of ordinary differential
equations with respect to \ $\tau$--variable. After some straightforward algebra
we obtain:
\begin{equation}
\beta p_2 = (1+\tau^2)^{3/2} {{\rm d} \psi \over {\rm d} \tau}  \sigma_0^{3/2}\Omega_0 ~,
\end{equation}
\begin{equation}
p_2 \rho_2 + { 2 \over 5} (1+\tau^2)^{3}
\left({{\rm d} \psi \over {\rm d} \tau}\right)^2
= { 2 \over 5} \frac{{\tau^2}}{1 + \tau^2}
\rho_2^2 \,\sigma_0^{3}\, \Omega_0^2 ~,
\end{equation}
where $\rho_2 \equiv \rho_{I2} \rho_{R2}$. Introducing notation:
$$ W \equiv {1 \over \rho_2} {{\rm d} \psi \over {\rm
d} \tau}
$$
we may derive the following algebraic equation:
\begin{equation}
\beta \left[ \beta W(\tau)^2 +
{5 \over 2} { \sigma_0^{3/2} \Omega_0 \over (1+\tau^2)^{3/2}} W(\tau) -
\beta \frac{\sigma_0^3 \Omega_0^2 \tau^2}{(1+\tau^2)^4} \right]  =  0 ~.
\label{Realizability}
\end{equation}
Eq. (\ref{Realizability}) links the values of the stream-function $\psi$,
$\tau$-dependent part of density $\rho_2$ and $\beta$ parameter
and represents the "realizability" condition for all existing solutions
within the considered Beltrami flow model of disk-jet structure formation.
There are three apparent solutions to this equation:
\begin{itemize}
\item[(i)] The solution with $\beta=0$, corresponding to the background
    dissipation model ($T_{ik} = \alpha_0 {\cal P}_0$); \vskip 0.3cm
\item[(ii)] Two separate solutions for the dissipative flow ($\beta \not=
    0$) with
\begin{equation}
W_{\pm}(\tau)  =
\label{Wpm}
\end{equation}
\[
- {5 \over 4}{\sigma_0^{3/2} \over {(1+\tau^2)^{3/2}}} {\Omega_0 \over \beta}
\left[ 1 \pm \left( 1 + {16 \over 25} {\beta^2 \tau^2  \over (1+\tau^2)} \right)^{1/2} \right]
.
\]
\vskip 0.3cm
\end{itemize}
For simplicity of the presentation we may constrain on the small $\beta$ limit
($\beta \ll 1$). Then the solutions given by (\ref{Wpm}) can be approximated by
the following simplified forms:
\begin{equation}
W_+(\tau) \approx {2 \over 5} {\tau^2 \over (1+\tau^2)^{5/2}} \beta \sigma_0^{3/2} \Omega_0 ~, \label{Wplus}
\end{equation}
\begin{equation}
W_-(\tau) \approx -{5 \over 2}{1\over (1+\tau^2)^{3/2}}
{\sigma_0^{3/2} \Omega_0 \over \beta} ~. \label{Wminus}
\end{equation}

\bigskip

Thus, the solutions of our disk-jet model can be calculated as follows
(where $W(\tau)$ is a general solution of (\ref{Realizability})):
\begin{eqnarray}
\rho(\sigma,\tau) &=& \sigma^{-3/2} \rho_2(\tau) \label{Solrho} ~ ,\\
p(\sigma,\tau) &=& {(1+\tau^2)^{3/2} \over \sigma^{5/2}}
{\sigma_0^{3/2} \Omega_0 \over \beta} \rho_2(\tau) W(\tau) ~ , \label{Sol1} \\
V_r(\sigma,\tau) &=& - \, {1+\tau^2 \over \sigma^{1/2}} W(\tau) ~ , \label{SolVr} \\
V_z(\sigma,\tau) &=& - \, \tau \,{1+\tau^2 \over \sigma^{1/2}} W(\tau) ~ . \label{Sol2}
\end{eqnarray}
Eqs. (\ref{Sol1}-\ref{Sol2}) together with radial profiles (41) and
appropriate choice of the solution for $W$ (see Eq. (\ref{Realizability}) give the full
solution for the disk-jet flow for different types of density profiles
$\rho_2(\tau)$. Hence, Eqs. (\ref{SolVr},\ref{Sol2}) indicate the character of
the solutions corresponding to the different sign of $W(\tau)$:
\begin{itemize}
\item Radial--vertical accretion flow for $W_+(\tau)>0$ corresponding to
    ~$V_r<0$ and $V_z<0$;
\item Radial--vertical ejection flow for $W_-(\tau)<0$ corresponding to
    $V_r>0$ and $V_z>0$;
\end{itemize}

\begin{figure}
\begin{center}
\includegraphics[width = 0.99 \columnwidth]{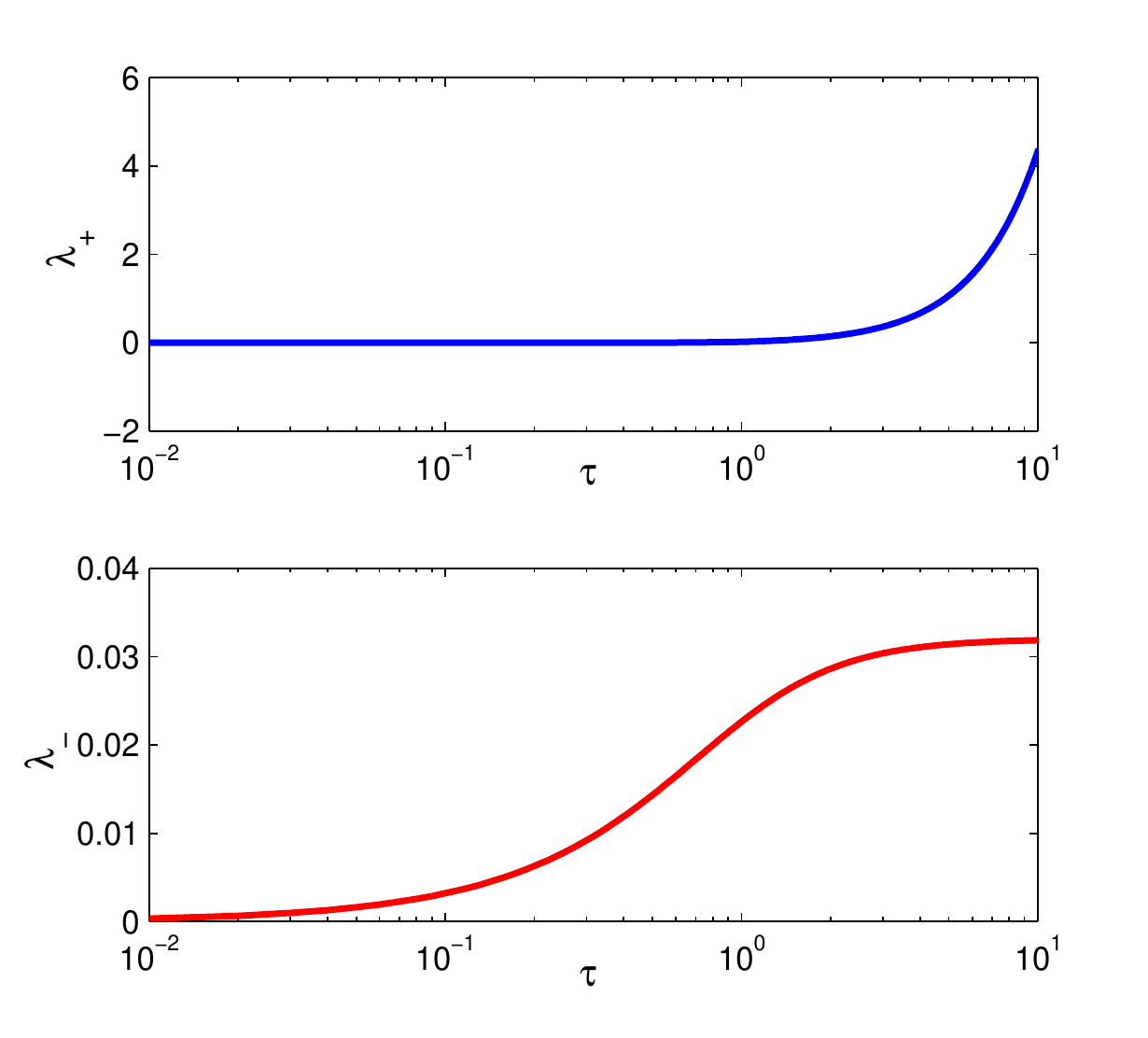}
\end{center}
\caption{The Beltrami parameters $\lambda(\tau)$ for
the disk (top) and the jet (bottom) solutions are shown vs $\tau$ for the case when
$\beta = 0.01$. Notice that Beltrami parameter for the disk solution is negligible
for low poloidal angles ($\tau \ll 1$). \label{lambda_tau}}
\end{figure}

Eqs. (\ref{Sim1},\ref{Sim4},\ref{Sim2}-\ref{Sim6}) allow us to calculate
dependence of the Beltrami parameter $\lambda$ on the turbulent viscosity
parameter  $\beta$ as follows:
\begin{equation}
\lambda = \left( {{\rm d} W \over {\rm d} \tau} + {5 \tau W \over 1 +\tau^2} \right)
\left({(1+\tau^2) W^2} + {\sigma_0^{3} \Omega_0^2 \over (1+\tau^2)^3 }\right)^{-1} \sigma_0^{3/2} \Omega_0
~.
\end{equation}
Interestingly, both solutions grow with $\beta$  (see Eqs.
(\ref{Wplus}) and (\ref{Wminus})):
$$
\lambda_{\pm} = \lambda(W_{\pm}) \propto \beta ~.
$$
Figure \ref{lambda_tau} shows solutions for the Beltrami parameters
\,$\lambda_{+}(\tau)$\, and \,$\lambda_{-}(\tau)$ \, corresponding to the
swirling flows that accrete and eject matter, respectively.

Substituting Eqs. (\ref{Wplus},\ref{Wminus}) into (\ref{SolVr}) and
(\ref{Sol2}) we may derive the velocity field components of the disk flow:
\begin{equation}
V_{rD}(\sigma,\tau)  = -{2 \over 5} \beta \sigma_0 \Omega_0
{\tau^2 \over (1 + \tau^2)^{3/2}}
\left( {\sigma \over \sigma_0} \right)^{-1/2} ~, \label{Sol3}
\end{equation}
\begin{equation}
V_{zD}(\sigma,\tau)  = -{2 \over 5} \beta \sigma_0 \Omega_0
{\tau^3 \over (1 + \tau^2)^{3/2}}
\left({\sigma \over \sigma_0}\right)^{-1/2} ~, \label{Sol4}
\end{equation}
and the jet flow:
\begin{equation}
V_{rJ}(\sigma,\tau) =  {5 \over 2} \frac{\sigma_0 \Omega_0}{\beta}
{1 \over (1 + \tau^2)^{1/2}}
\left( {\sigma \over \sigma_0} \right)^{-1/2} ~, \label{VrJ}
\end{equation}
\begin{equation}
V_{zJ}(\sigma,\tau)  =  {5 \over 2} \frac{\sigma_0 \Omega_0}{\beta}
{\tau \over (1 + \tau^2)^{1/2}}
\left( {\sigma \over \sigma_0} \right)^{-1/2} ~. \label{VzJ}
\end{equation}

\begin{figure}
\begin{center}\includegraphics[width= 0.95 \linewidth]{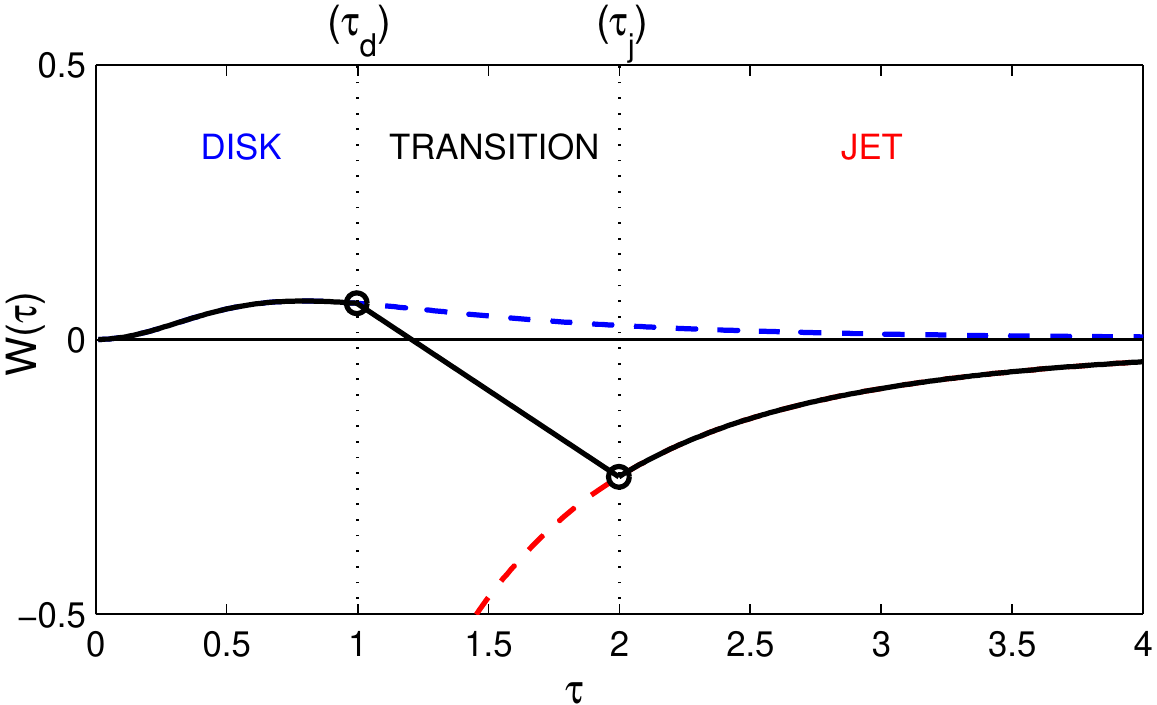}\end{center}
\caption{$W(\tau)$ (solid black line) vs $\tau$ in three region solution.
$W_+(\tau)$ and $W_-(\tau)$ are shown by blue and red dashed lines, respectively.
$W(\tau)$ follows $W_+$ in the disk region and $W_-$ in the jet region.
Transition region in this models starts at $\tau_d=1$ and ends at $\tau_j=2$.
Within the transition region, where flow is effectively ballistic, disk
solution can continuously switch into the jet solution.} \label{W3}
\end{figure}

\subsection{Disk-jet structure solutions}
\label{Structure}

Analysis present above show that Beltrami flow model for disk-jet structure
formation with turbulent viscosity assumption is able to describe classes of
solutions corresponding to the accretion disk--ejection jet flow
(one flow with the matter and energy). In this model kinematics of the
solution can be derived using the realizability parameter $W(\tau)$ that sets
the topology of the solution. To get disk-jet configuration in one solution we
need to construct flow that matches the disk solution with $W_+$ at $\tau \ll
1$ and jet solution for nearly vertical direction ($\tau \gg 1$).
The requirement of the continuity of the velocity field
restricts any jumps in $W(\tau)$. Hence, to describe the disk-jet solution we
use three region model:
\begin{itemize}
\item[1.)] Disk region for $\tau < \tau_d$ when $W(\tau) = W_+(\tau)$ \ ,
    \vskip 0.2cm
\item[2.)] Transition region for $\tau_d < \tau < \tau_j$ when
    $\beta=0$ \ , \vskip 0.2cm
\item[3.)] Jet region for $\tau > \tau_j$ when $W(\tau) =
    W_-(\tau)$ \ , \vskip 0.2cm
\end{itemize}
where  $\tau_d$ and  $\tau_j$ are the maximal and minimal values of $\tau$ in
the disk and jet regions, respectively. We assume that closer to the central
object flow passes through a transition region, where it goes through a
ballistic regime, hence, undergoing the dramatic acceleration in vertical
($z$) direction -- jet launching. In this region $\beta=0$ and "realizability"
condition allows any values of $W(\tau)$. Hence, $W_+(\tau_d)$ can switch into
$W_-(\tau_j)$ {\it continuously} within the transition region. Figure \ref{W3}
illustrates the model of $W(\tau)$ for the proposed three region solution.
In this model a viscous accreting flow in the disk region
($\tau < \tau_d$) goes through inviscid ballistic regime in the transition
region ($\tau_d < \tau < \tau_j$) and, finally, into the viscous outflow configuration
in the jet region ($\tau > \tau_j$). Hence, the continuous velocity field of
this model can be calculated using Eqs. (\ref{SolVr}), (\ref{Sol2}) with
the three region model of the $W(\tau)$ function constructed
above.

\begin{figure}
\begin{center}\includegraphics[width= 0.98 \linewidth]{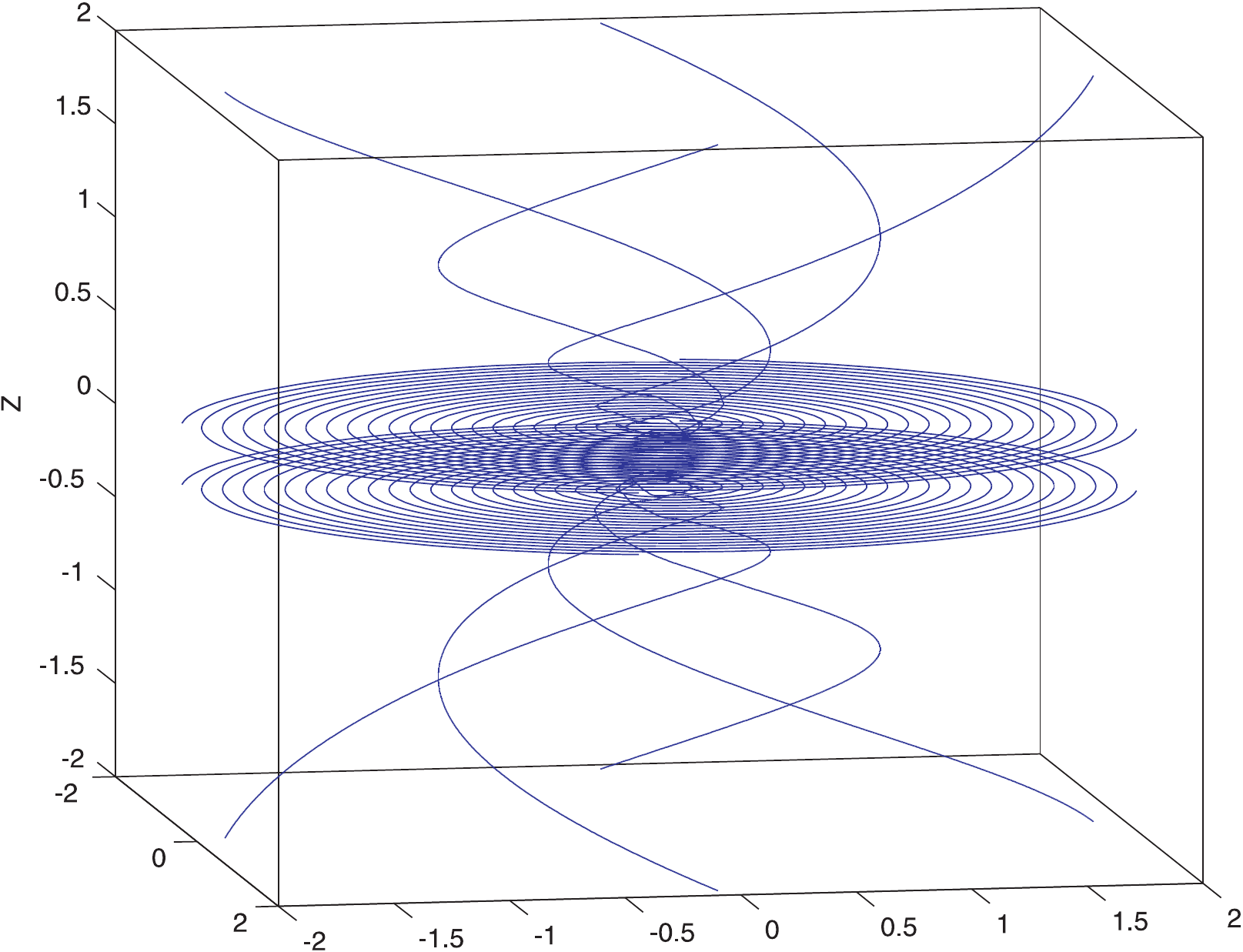}\end{center}
\caption{Velocity streamlines of the disk-jet structure illustrating
accretion-ejection flow at $\tau_d=1$, $\tau_j=2$, $\tau_0=0.01$ $\beta=0.02$.
Decrease of the $\beta$ parameter leads to the increase of the
ratio between the vertical and radial velocities and, consequently, change of the flow geometry.}
\label{streamline}
\end{figure}

Then we can estimate the accretion speed of the flow in the disk region ($\tau
< \tau_d$):
\begin{equation}
V_{\rm acc} = \left(V_{rD}^2 + V_{zD}^2 \right)^{1/2} =
{2 \over 5} {\tau^2 \over (1+\tau^2)^{1/2}} ~ \beta V_{\rm Kep} ~, \label{Vacc}
\end{equation}
and ejection velocity in the jet region ($\tau > \tau_j$):
\begin{equation}
V_{\rm ej} = \left(V_{rJ}^2 + V_{zJ}^2 \right)^{1/2} =
{5 \over 2} \left( 1+\tau^2 \right)^{1/2} ~ {V_{\rm Kep} \over \beta} ~. \label{Vej}
\end{equation}
Thus, in the low $\beta$ limit derived solution corresponds to the locally
slowly accreting flow ($V_{\rm acc} \ll V_{\rm Kep}$) with the
locally fast outflow in the jet region ($V_{\rm ej} \gg V_{\rm Kep}$),
matching the properties of astrophysical accretion-ejection flows. Notice, that
above expressions for local accretion flows in the disk and local outflows
in the jet do not depend on the explicit profile of $\tau $-dependent part of density.
Figure \ref{streamline} shows the velocity streamlines of the derived disk-jet structure.

\begin{figure}
\begin{center}
\includegraphics[width=1.0 \columnwidth]{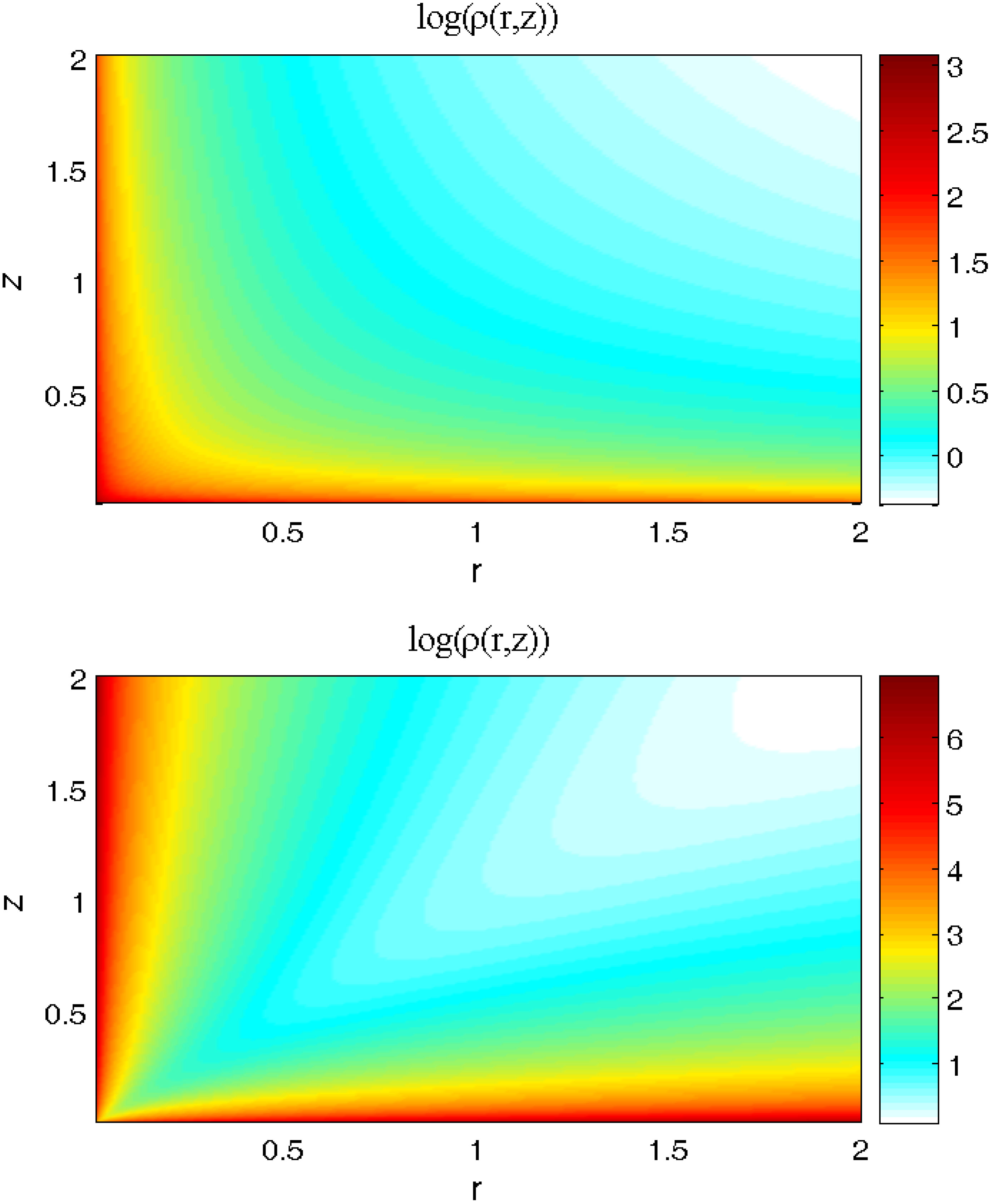}
\end{center}
\caption{Total density $\rho(r,z)$ distribution of the disk-jet structure is shown for:
$A_d=1$, $A_j=1$, $m_d=-1$, $m_j=1$ (top) and
$A_d=3$, $A_j=3$, $m_d=-3$, $m_j=3$ (bottom). n all cases $\tau_0$ = 0.01.
The topology of the density distribution is set by the disk ($m_d<0$) and the jet
$(m_j>0)$ power indices; for $\rho_2(\tau)$ the power-law
distribution (\ref{Rho2SY}) was used .} \label{surfs}
\end{figure}

To close the system of the disk-jet solutions we need to define angular
distribution of the density $\rho_2(\tau)$ that will be used to obtain
different classes of disk-jet structures through Eqs.
(\ref{Solrho}-\ref{Sol2}). For this purpose we employ the power-law
distribution (cf. \citealt{SY2011}):
\begin{equation}
\rho_2(\tau) = A_{\rm d} (\tau + \tau_0)^{m_{\rm d}} +
A_{\rm j} (\tau + \tau_0)^{m_{\rm j}} ~, \label{Rho2SY}
\end{equation}
where parameters $A_d$, $m_d$ and $A_j$, $m_j$ define the density profile in
the disk and jet regions, respectively. The small parameter $\tau_0 \ll 1$ is
used to avoid divergence at the disk center $\tau=0$. Hence, Eqs.
(\ref{Solrho}-\ref{Sol2}) and ({\ref{Rho2SY}}) describe classes of disk-jet
solutions within our self-similar Beltrami flow model. Figure \ref{surfs} shows
the density distribution of the disk-jet structure.

\subsection{Properties of the disk-jet structure}
\label{Properties}

The purpose of the current paper is to find the analytical
solutions constructing a reliable modal for disk-jet structure formation that
describe basic properties of hydrodynamic jet outflows from YSOs. Solutions
derived in the paper depend on number of parameters. Below we evaluate the
possibility of linking these parameters with observational properties of YSOs.

Eqs. (\ref{Vacc}) and (\ref{Vej}) reveal the link of the $\beta$ parameter
with kinematic properties of disk-jet structures:
\begin{equation}
\beta^2 = {V_{\rm acc} \over V_{\rm ej}} ~. \label{beta2}
\end{equation}
The value of $\beta$ parameter is constrained by the $\alpha_0$ parameter
that describes anomalous viscosity due to background stationary turbulence
(see Eqs. (\ref{T0ik}, \ref{tik})): $\beta < \alpha_0$. Using a typical value
from observational luminosity $\alpha_0 \sim 0.01$ we can get:
\begin{equation}
V_{\rm ej} > 10^{4} V_{\rm acc} ~.
\end{equation}
Specific value of the $\beta$ parameter can be inferred from observations,
where both the radial accretion and the vertical ejection velocities near the
central object can be observed.

\bigskip

\begin{figure}
\begin{center}
\includegraphics[width=0.95 \columnwidth]{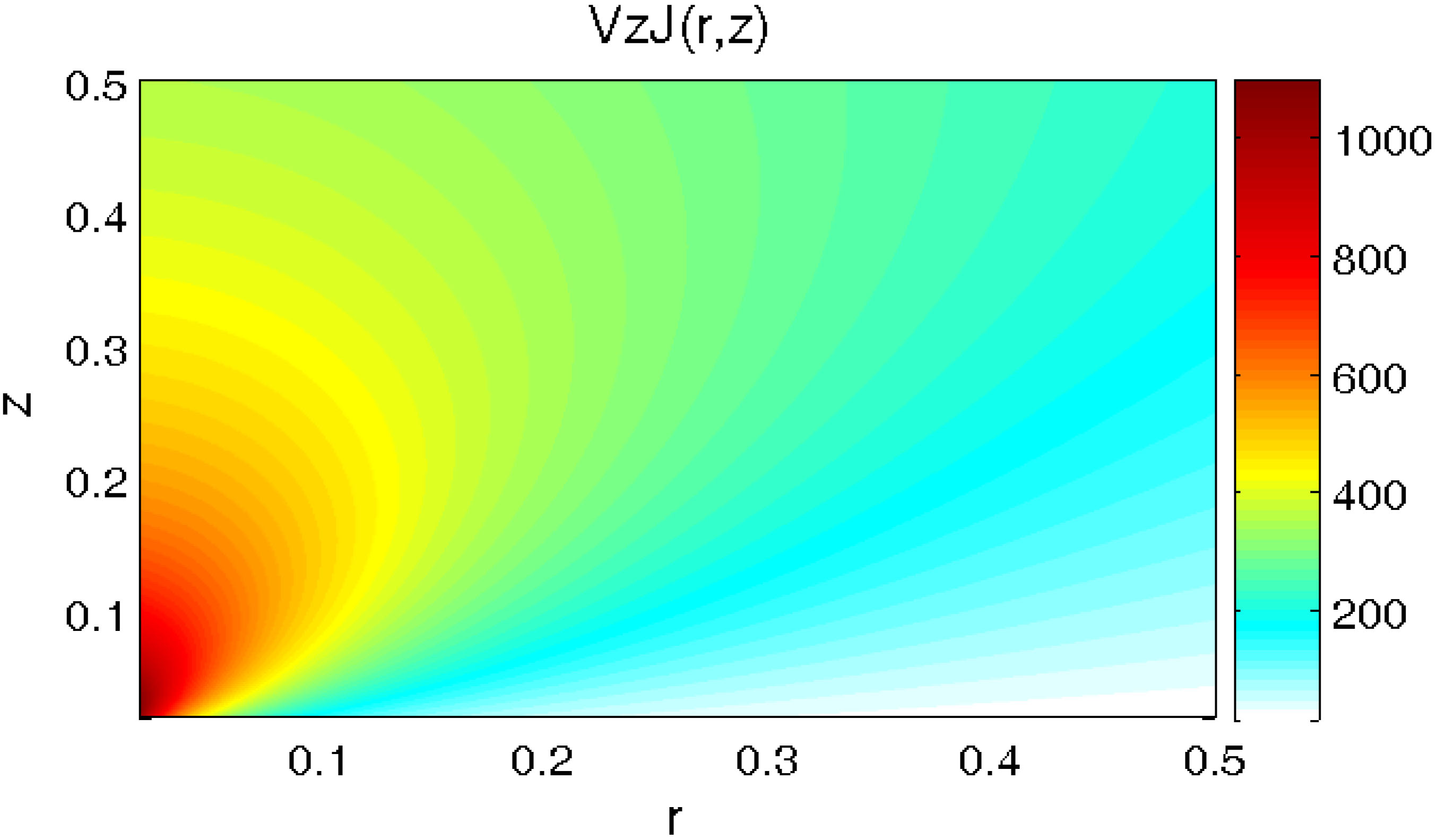}
\end{center}
\caption{Vertical jet velocity of the jet solution
$V_{zJ}(r,z) / (\sigma_0 \Omega_0)$ (see Eq. (\ref{VzJ})).
Here $\beta = 0.01$ and $\sigma_0=1$.
Maximal velocity of the outflow is reached near the vertical axis above
the disk plane, at the top edge of transition region. \label{vzj}}
\end{figure}

One of the major properties/features of the astrophysical disk-jet flows is
their narrow high velocity vertical jets. To illustrate the outflow properties
of our solutions the vertical velocity distribution of the jet flow is plotted
in Figure \ref{vzj} (see Eq. (\ref{VzJ})). The outflow velocity is maximal at
the top edge of transition region, above the disk-plane; beyond this maximum
the vertical flow velocity decreases both with vertical and radial distances,
similar to the Keplerian profile ($\propto (\sigma / \sigma_0)^{-1/2}$);
the outflow launching is at the bottom edge of transition region, just above
the disk-surface. Hence, this means that solutions derived within
the minimal Beltrami flow model can describe the {\it formation} of the
disk-jet structure (Shatashvili \& Yoshida 2011), while the effects of the
formed jet acceleration and collimation occur in the vertical outflows farther
out from the central object and requires more general dynamical model
including the heating/cooling processes.

\bigskip

\begin{figure}
\begin{center}
\includegraphics[width=0.95 \columnwidth]{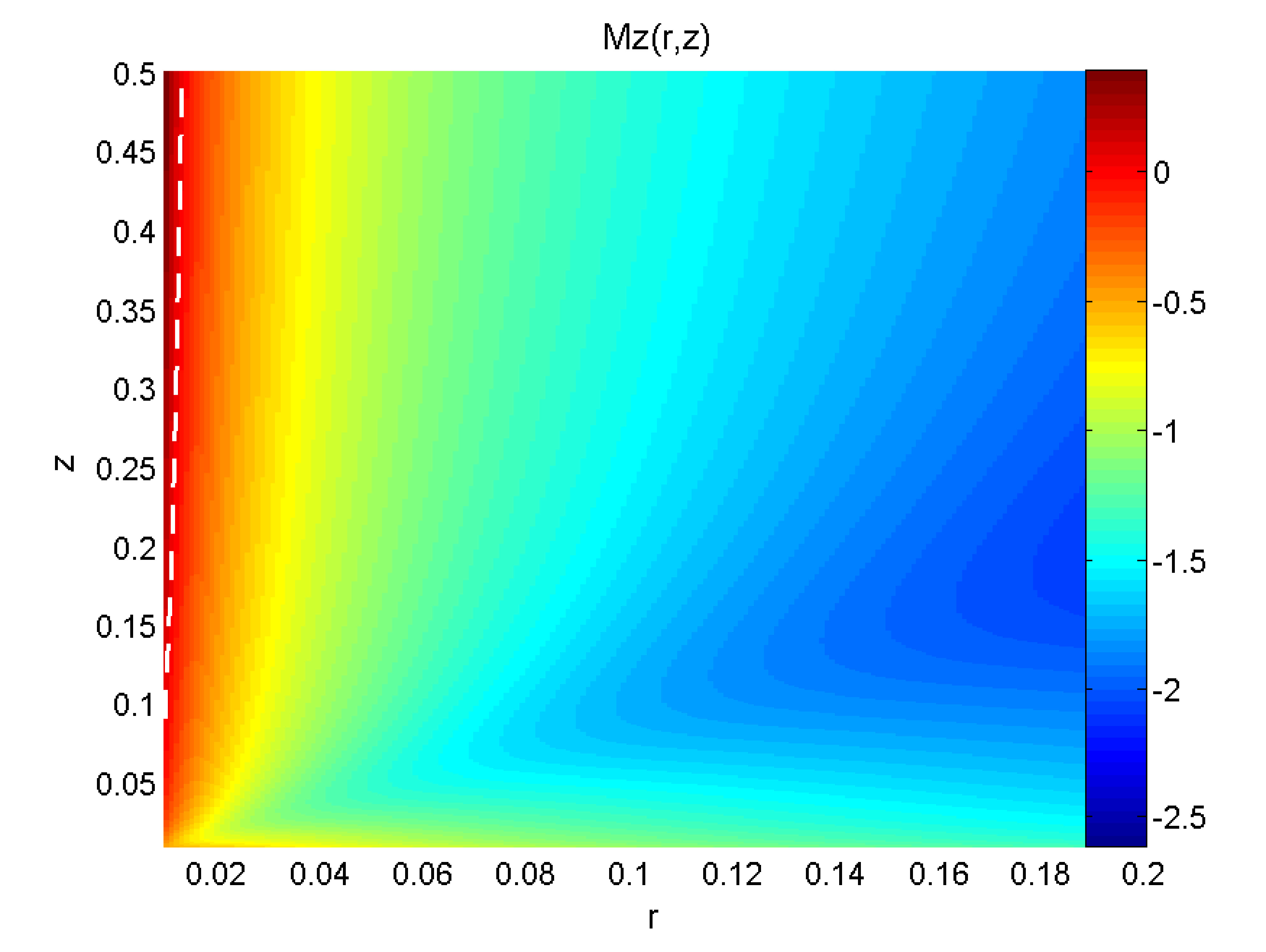}
\end{center}
\caption{Vertical Mach number  of the jet flow ${\rm log}(M_z(r,z))$ when
$A_d=3$, $A_j=3$, $m_d=-3$, $m_j=3$. $\tau_0=0.01$, $\beta = 0.01$,
$\sigma_0=1$ and ${\cal P}_0/p_0 = 10^{6}$. Vertical dashed line shows area, where
the Mach number reveals supersonic flow: $M_z>1$. For density the distribution
presented in Figure \ref{surfs} was used.}
\label{Mz}
\end{figure}

The standard mechanism of the jet acceleration through purely hydrodynamic
mechanism is the Laval nozzle, when adiabatic expansion of the supersonic flow
leads to its acceleration. To estimate the feasibility of this mechanism for
the derived in this paper disk-jet structure we calculate the local Mach
number of the vertical jet flow defined as follows:
\begin{equation}
M_z = {V_{zJ} \over
[({\cal P}_0 + p) / \rho]^{1/2}}~.
\end{equation}
Note that pressure variation induced by disk-jet solutions is negative in the
jet region ($p<0$ when $W_-(\tau)<0$). Hence, swirling solution in the jet
region leads to the decrease of the pressure and corresponding sound speed,
thus increasing the local Mach number of the flow. Background pressure can be
normalized on the pressure of the self-similar solution at $\sigma=\sigma_0$ and
$\tau = \tau_j$ (see Eqs. (\ref{Sol1}, \ref{Rho2SY})):
\begin{equation}
p_0 \approx {5 \over 2} {\sigma_0^{1/2} \Omega_0^2 \over \beta^2} A_j \tau_j^{m_j} ~,
\end{equation}
assuming that $\tau_0 \ll \tau_j$.  Figure \ref{Mz} shows local Mach number of
the jet outflow for ${\cal P}_0/p_0=10^{5}$. It seems that the decrease of the
vertical outflow velocity coincides with simultaneous decrease of the sound
speed, thus, leading to the possible increase of the local Mach number. In the
considered extreme limit the flow can reach supersonic velocities in the
narrow jet region at $\tau \gg 1$. The increase of the
background pressure would decrease the local Mach number to the subsonic
values. At wider polar angles ($\tau>1$) vertical flow is
subsonic and thus should be decelerating away from the central object (see
also Figure \ref{vzj}):
$$
V_z(\sigma) \propto \sigma^{-1/2} ~.
$$
Indeed, such wide-angle rotating outflows near the central object together
with narrow collimated jets are consistent with ALMA observations of the HH
objects (see \cite{Acre2013}). Moreover, the outer parabolic shape of the
wide-angle outflow near the central object seen in color gradients
illustrated in Figure \ref{Mz} is observed
recently for the Class $0$ protostellar system by \cite{Lee2018}.

In realistic disk-jet systems jet flow streaming away from the central object
is likely to undergo cooling (effect not considered in our model) that will
further reduce the local sound speed and may render the ourflow velocity to
become supersonic. In this case farther adiabatic expansion will lead to the
jet acceleration -- property inherent to the protostellar disk outflows. Thus,
solution derived in present study using Generalized Beltrami flow
configuration for disk-jet structure can
successfully mimic slow radial sub-Keplerian accretion flow in the disk region
and fast narrow super-Keplerian outflow in the jet region. The decrease of the
$\beta $ parameter leads to the increase of the ratio between the vertical and
radial velocities and, consequently, change of the disk-jet flow geometry.

\section{Summary}
\label{Summary}

We study the hydrodynamic disk-jet structure formation phenomenon for YSOs
based on the analytic Beltrami-Bernoulli model using the extended turbulent
viscosity assumption as the main reason of the accretion in the disk. For this
purpose we consider the stationary turbulent state powering the accretion
process and leading to the collimated jet outflow.

We have employed generalized turbulent viscosity model to describe the effect
of turbulent dissipation in both, disk as well as the jet areas
of a disk-jet structure. For this
purpose we have split the turbulent viscosity tensor into the constant
background and varying deviation from the constant profile, leading to the
analytic formulation of the viscosity effects in the disk-jet transition area.
In this formalism we factorize the local physical parameters of flow into the
``ideal fluid'' and ``reduced'' components following \citealt{SY2011}
Beltrami-Bernoulli model; this allows to find the self-similar solutions of
the disk-jet flow in the field of central gravitating object.

We have formulated the realizability condition for the solutions
that reveal the class of disk-jet flow solutions smoothly distributed in
three well-defined connected regions of a global structure:
a) flow accreting in the radial and vertical direction,
b) flow ejecting in the radial and vertical direction; c)
flow in the ballistic regieme. Hence, constructing the global solution using
the disk inflow at low poloidal angles and jet outflow at high poloidal
angles, having ballistic transition from one to another, we have derived
disk-jet structure with slow accretion and high ejection velocities.

It seems that our disk-jet structure depends on the thermal properties of the
disk flow.

Local Mach number of the outflow depends on the background pressure in the jet
area. At low pressure, i.e., low temperature values jet outflow may reach
supersonic amplitudes close to the central axis of the flow at high poloidal
angles. At lower poloidal angles outflow is subsonic, thus
showing wide-angle outflow decelerating with radial distance and decrease of
the polidal angle. Considered disk-jet solution describes the formation of
the high velocity outflow from slowly accreting Keplerian
disk-flow. The further kinematic acceleration and collimation of the jet flow
may be due to the effects not considered in the current minimal model. In case
of YSOs, our solution shows weaker jets at the later stage of the evolution of
protostar, when the temperature of the central object and corresponding disk
matter increases.

\bigskip

In our analysis magnetized disk-winds were not invoked
-- disk was assumed to be not ionized (hence, magnetic fields can not
affect the flow structure). We believe, that additional effects of magnetic fields (self-consistently
generated or advected, or their combination) will make the solution of the
problem only richer; the consequent problems of jet acceleration and heating
could be also discussed then. Due to Hall effect the generalized
magneto-Bernoulli mechanism (\citealt{mnsy,MSaccel}) may effectively
accelerate the jet-flow. In a weakly ionized plasma the Hall effect is
magnified by the ratio of the neutral and electron densities; the ambipoler
diffusion effect also yields a higher-order perturbation
(\citealt{krishan2006}). The disk-jet connection point may differ depending on
the plasma condition near the central object: e.g. in AGN, the plasma is fully
ionized but rather collisional (we also need a relativistic equation of state
with possible presence of pairs). Also the details of dissipation mechanism
may play the additional role. Such issues will be discussed in our future
works.

Analytic solutions for disk-jet structure derived in the present paper can be used to analyze the
properties of hydrodynamic jets from YSOs, and link their topological
properties to the local physical conditions at the jet launching areas.

Our derived Disk-jet solutions describe the astrophysical
disk-jet structures with low ionization, where the main energy source of the
outflow should come from non-magnetic processes. These should include
hydrodynamic jets from protostellar accretion disks and young stellar objects
in general. Our analytic model links the accretion and ejection rates of the
disk-jet flow, thus allowing to propose the specific predictions for observed
structures.

\section*{Acknowledgments}
\label{Ackno}

Authors express special thanks to Prof. Zensho
Yoshida for his valuable discussions. E.A., B.M., M.G., V.L. and I.J.
acknowledge support from the TSU Student Research Council and the TSU Faculty
of Exact and Natural Sciences Students Grant; M.G.-s and I.J.-s work was
partially supported by World Federation Of Scientists National Scholarship
Programme Geneva, 2018; NLS-s, M.G.-s and I.J.-s research was partially
supported by Shota Rustaveli Georgian National Foundation Grant Project No.
FR17-391.




\begin{thebibliography}{00}


\bibitem[ ()]{}

\bibitem[Arce et al. (2013)]{Acre2013}
    Arce, H. G., Mardones, D., Corder, S. A., Garay, G., Noriega-Crespo, A.,
    Raga, A. C., 2013
    ApJ, 774, 39

\bibitem[Anderson et al. (2005)]{anderson2}
    Anderson, J. M., Li, Z. Y.,  Krasnopolsky, R. and Blandford, R. D., 2005
    ApJ, 630, 945

\bibitem[Bally (2016)]{Bally2016}
    Bally, J., 2016
    Ann. Rev. Astron. Astrophys., 54, 491

\bibitem[Begelman et al. (1984)]{begelman3}
    Begelman, M. C., Blandford, R. D. and Rees, M. J., 1984
    Rev. Mod. Phys. 56, 255

\bibitem[Begelman (1993)]{begelman4}
    Begelman, M. C., 1993
    ''Conference summary'', in {\it Astrophysical
    Jets}, \ ed. D. Burgarella et al (Cambridge: Cambridge Univ.
    Press), 1993, pp. 305-315.

\bibitem[Begelman (1998)]{begelman5}
    Begelman, M. C., 1998
    ApJ, 493, 291

\bibitem[Belan et al. (2013)]{Belan2013}
    Belan M., Massaglia S., Tordella D., Mirzaei M., and de Ponte S.,
    2013, A\&A, 554, A99

\bibitem[Bisnovatyi-Kogan \& Lovelace (2007)]{lovelace2}
    Bisnovatyi-Kogan G. S. and Lovelace, R.V.E., 2007,
    ApJ 667(2), L167-L169

\bibitem[Blandford \& Rees (1974)]{bland}
   Blandford,  R. D. and Rees, M. J., 1974
   MNRAS, 169, 395

\bibitem[Blandford \& Znajek (1997)]{bland1}
    Blandford, R. D. and Znajek, R. L., 1977
    MNRAS, 179, 433

\bibitem[Blandford \& Payne (1982)]{bland1-2}
    Blandford, R. D. and Payne, D. G., 1982
    MNRAS, 199, 883

\bibitem[Blandford (1994)]{bland2}
    Blandford, R. D., 1994
    ApJS, 90, 515

\bibitem[Bjerkeli et al. (2016)]{Bjerkeli2016}
    Bjerkeli, P., van der Wiel, M. H. D., Harsono, D., Ramsey, J. P.,
    Jorgensen, J. K., 2016, Nature, 540, 406

\bibitem[Celloti \& Blandford (2001)]{bland4}
    Celotti, A.  and Blandford, R. D., ''Black Holes in Binaries and
    Galactic Nuclei: Diagnostics, Demography and Formation'', in {\it
    ESO Astrophysics Symposia} \  ed. L. Kaper \textit{et al.}
    (Berlin, Heidelberg: Springer-Verlag), 2001, 206.

\bibitem[Clarke \& Alexander (2016)]{Clarke2016}
    Clarke, C. J., and Alexander, R. D., 2016
    MNRAS, 460, 3044

\bibitem[Fereira (1996)]{Fereira1997}
    Ferreira, J., 1997,
    A\&A 319, 340

\bibitem[Fereira et al. (2006,2007)]{Fereira}
    Ferreira, J., Dougados, C. and Cabrit, S., 2006
    A\&A 453, 785 (2006);
    Ferreira, J., Dougados, C. and Whelan, E.,
    ”Jets from Young Stars I: Models and Constraints” in Lecture Notes in Physics
    ed. Ferreira, J.  et al. (Berlin, Heidelberg: Springer-Verlag) 2007, 723, 181.

\bibitem[Hartigan et al. (2011)]{Hartigan2011}
    Hartigan P., Frank A., Foster J. M., Wilde B. H., Douglas M., Rosen P. A.,
    Coker R. F., Blue B. E., and Hansen J. F., 2011, ApJ, 736, 29

\bibitem[Hernandez et al. (2014)]{Hernandez2014}
    Hernandez, X., Rendon, P. L., Rodriguez-Mota, R. G., and Capella A., 2014
    Rev. Mex. Astr. Astrophys. 50, 23.

\bibitem[Ioannidis \& Froebrich (2012)]{Ioannidis2012}
    Ioannidis, G., Froebrich, D., 2012,
    MNRAS, 421, 3257

\bibitem[Jhan \& Lee (2016)]{Jhan2016}
    Jhan, K.-S., Lee, C.-F., 2016, ApJ, 816, 1

\bibitem[Krasnopolsky et al. (1999)]{bland3}
    Krasnopolsky, R., Li, Z. Y.  and Blandford, R. D., 1999
    ApJ, 526, 631

\bibitem[Krishan \& Yoshida (2006)]{krishan2006}
    Krishan V., Yoshida Z., 2006
    Phys. Plasmas, {13}, 092303


\bibitem[Kudoh \& Shibata (1997)]{shibata}
    Kudoh, T.  and Shibata, K., 1987
    ApJ, 474, 362

\bibitem[Kuwabara et al. (2005)]{shibata2}
    Kuwabara, T., Shibata, K., Kudoh,  T. and Matsumoto, R., 2005
    ApJ, 621, 921

\bibitem[Lee et al. (2017)]{Lee2017}
    Lee, C.-F., Ho, P. T., Li, Z.-Y., Hirano, N., Zhang, Q., Hsien, S., 2017,
    Nature Astr., 1, 0152

\bibitem[Lee et al. (2018)]{Lee2018}
    Lee, C.-F., Li, Z.-Y., Codella, C., Ho, P. T. P., Podio, L., Hirano, N.,
    Shang, H., Turner, N. J., Zhang, Q., 2018
    ApJ, 856, 14


\bibitem[Livio (1997)]{Livio}
    Livio, M. ”The Formation Of Astrophysical Jets",
    in Accretion Phenomena and Related Outflows;
    IAU Colloquium 163 ed. D. T. Wickramasinghe et al
    (San Francisco: ASP) ASP Conference Series 1997, 121, 845.

\bibitem[Lizano (1988)]{Lizano1988}
    Lizano, S., Heiles, C., Rodriguez, L. F., Koo, B.-C., Shu, F. H.,
    Hasegawa, T., Hayashi, S., Mirabel, I. F., 1988
    ApJ, 328, 763


\bibitem[Lovelace et al. (1994)]{lovelace1}
    Lovelace, R.V.E., Romanova, M.M. and Newman, W.I., 1994,
    ApJ, 437, 136

\bibitem[Lubow et al. (1994)]{lubow}
    Lubow, S.H., Papaloizou, J.C.B. and Pringle, J.E., 1994,
    MNRAS 267, 235

\bibitem[Mahajan et al. (2002)]{mnsy}
    Mahajan S. M., Nikol'skaya K. I., Shatashvili N. L., Yoshida Z., 2002
    ApJ, 576, L161

\bibitem[Mahajan et al. (2006)]{MSaccel}
    Mahajan S. M., Shatashvili N. L., Mikeladze S. V.,  Sigua K. I., 2006
    Phys. Plasmas, 13, 062902

\bibitem[Moss et al. (2013)]{Moss2013}
    Moss, V. A., McClure-Griffiths, N. M., Murphy, T., Pisano, D. J.,
    Kummerfeld, J. K., Curran, J. R., 2013, ApJS, 209, 12

\bibitem[Plunkett et al. (2015)]{Plunkett2015}
    Plunkett, A. L., Arce, H. G., Mardones, D., van Dokkum, P., Dunham, M.
    M., Fernandez-Lopez, M., Gallardo, J., Corder, S. A., 2015,
    Nature, 527, 70

\bibitem[Podio et al. (2016)]{Podio2016}
    Podio, L., Codella, C., Gueth, F., Cabrit, S., Maury, A., Tabone, B.,
    Lefevre, C., Anderl, S., Andre, P., Belloche, A., Bontemps, S.,
    Hennebelle, P., Lefloch, B., Maret, S., Testi, L., 2016,
    A\&A, 593, L4

\bibitem[Reiter et al. (2017)]{Reitar2017}
    Reiter, M., Kiminki, M. M., Smith, N., Bally, J., 2017,
    MNRAS, 467, 4441

\bibitem[Ruden et al. (1990)]{Ruden1990}
    Ruden, S. P., Glassgold, A., and Shu, F., 1990
    ApJ, 361, 546

\bibitem[Scott \& Lovelace (1987)]{Scott1987}
    Scott, H. A. and Lovelace, R. V. E., 1987,
    Ap. \& SS, 129, 361

\bibitem[Shu et al. (1991)]{Shu1991}
    Shu, F., Ruden, S. P., Lada, C. J., Lizano, S., 1991
    ApJ, 370, L31

\bibitem[Squire (1951)]{Squire1951}
    Squire, H. B., 1951
    Quart. J. Mech. Appl. Math., 4, 321

\bibitem[Shakura \& Sunyaev (1973)]{Shakura1973}
    Shakura, N. I., Sunyaev, R. A., 1973,
    A\&A, 24, 337

\bibitem[Shatashvili \& Yoshida (2011)]{SY2011}
    Shatashvili, N.L. and Yoshida, Z., 2011,
    AIPCP, 1445, 34-53

\bibitem[Smith et al. (2014)]{Smith2014}
    Smith, M. D., Davis, C. J., Rowles, J. H., Knight, M., 2014,
    MNRAS, 443, 2612

\bibitem[Tordella et al. (2011)]{Tordella2011}
    Tordella, D., Belan, M., Massaglia, S., De Ponte, S., Mignone, A.,
    Bodenschatz, E., Ferrari, A., 2011,
    New J. Phys., 13, 043011

\bibitem[Yoshida \& Shatashvili (2012)]{Yoshida2012}
    Yoshida, Z., Shatashvili N. L., 2012,
    arXiv:1210.3558

\bibitem[Zanni et al. (2007)]{zanni}
    Zanni, C., Ferrari, A., Rosner, R., Bodo,  G. and Massaglia, S., 2007
    A\&A, 469, 811

\bibitem[Zhang et al. (2014)]{Zhang2014}
    Zhang, M., Wang, H., Hennings, T., 2014,
    AJ, 148, 26

\end{thebibliography}



\end{document}